\documentclass[twocolumn,showpacs,preprintnumbers,prl,amsmath,amssymb]{revtex4}
\usepackage{graphicx}
\usepackage{amssymb}
\usepackage{bm}
\usepackage{bbold}

\begin{document}

\title{Cluster update for tensor network states}

\author{Ling Wang}
\affiliation{Faculty of physics, Boltzmanngasse 5, 1090 Vienna, Austria}
\author{Frank Verstraete}
\affiliation{Faculty of physics, Boltzmanngasse 5, 1090 Vienna, Austria}
\date{\today}

\begin{abstract}
  We propose a novel recursive way of updating the tensors in
  projected entangled pair states by evolving the tensor in imaginary
  time evolution on clusters of different sizes. This generalizes the
  so-called simple update method of Jiang et
  al. [Phys. Rev. Lett. 101, 090603 (2008)] and the updating schemes
  in the single layer picture of Pi\v{z}orn et al. [Phys. Rev. A 83,
  052321 (2011)]. A finite-size scaling of the observables as a
  function of the cluster size provides a remarkable improvement in
  the accuracy as compared to the simple update scheme. We benchmark
  our results on the hand of the spin 1/2 staggered dimerized
  antiferromagnetic model on the square lattice, and accurate results
  for the magnetization and the critical exponents are determined.
\end{abstract}

\maketitle
Numerical simulation of strongly correlated quantum systems in
dimension above 1 remains one of the big challenges in condensed
matter physics. Quantum monte carlo fails for models suffering from
the sign problem, and the density matrix renormalization group method
(DMRG) has a problem in 2D because the violation of the area law of
entanglement entropy~\cite{valentinadvphys}. Tensor network states
(TNSs)~\cite{nishino-pre64-016705,frank04,vidal08}, on the other hand,
naturally generalize MPSs to higher dimensions and fulfill the area
law of entanglement entropy, and are proving to provide powerful tools
to simulate problems in above 1
dimension~\cite{murg,orus,jordan,evenbly,corboz,gu,zhou,xiang,su,kao,sandvik,fermiontns}.

One of the major difficulties in a TNS algorithm is the scaling of the
computational demands as a function of the virtual bond dimension $D$.
Inspired by the DMRG algorithm, where truncation is made with respect
to the reduced density matrix, a contraction algorithm at the
wavefunction level was recently proposed ~\cite{iztok}. However, the
rapidly growing renormalized physical index with the system size
causes a barrier that hampers the efficiency of the algorithm. The
simple update~\cite{xiang08}, proposed as a generalization of the
infinite time-evolving block decimation (iTEBD)
algorithm~\cite{vidal07} to TNSs, successfully avoids the
exponentially large Hilbert space, and is very
efficient~\cite{xiang,su,ling,linprb84}. However, the product
environment is too simple to capture the long range entanglement and
correlations near the critical point of a second order phase
transition~\cite{linprb84,su}. The efficiency of the simple update
still sheds light at controlling the Hilbert space in the complete
contraction algorithm~\cite{iztok}. The goal of this paper is to
demonstrate that those methods can be merged together in such a way
that the advantages of both methods are preserved.

\begin{figure}
\begin{center}
\includegraphics[width=8.5cm]{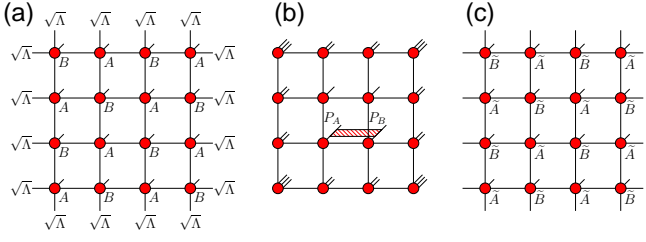}
\caption{(a) Wavefunction of an infinite lattice in terms of
  $l_x\times l_y$ local tensors and the entanglement spectra placed on
  each open bond. (b) An equivalent OB PEPS to (a), with 2 sites acted
  by evolution operator in the center of the cluster. (c) After
  projection with projectors $\mathbf{P}_A,\mathbf{P}_B$, and replace
  $\mathbf{A,B}$ with $\widetilde{\mathbf{A}}, \widetilde{\mathbf{B}}$
  everywhere, a new iPEPS wavefunction is obtained.}
\label{wavefunc1}
\end{center}
\end{figure}

{\it Method} -- The wavefunction of an infinite projected entangled
pair state (iPEPS) in terms of local tensors on 2 sites in the simple
update~\cite{xiang08} is given by
\begin{eqnarray}
\nonumber
|\psi\rangle&=&\sum_{s_a,s_b}\sum_{ijklmnp} (\Lambda_{ii}\Lambda_{jj}\Lambda_{kk}\Lambda_{mm}\Lambda_{nn}\Lambda_{pp})^{\frac{1}{2}}\\
&&\times T^{s_a}_{ijkl}T^{s_b}_{lmnp}|s_a\rangle\otimes |s_b\rangle \otimes |s_E\rangle,
\end{eqnarray}
where $|s_E\rangle=|i,j,k,m,n,p\rangle$ represents the environmental
degrees of freedom. We generalize this construction to a block of
$l_x\times l_y$ sites with the open virtual bonds weighted by the
diagonal tensors $\sqrt{\mathbf{\Lambda}}$s, where $\mathbf{\Lambda}$s
are the entanglement spectra with respect to each bond, as in
Fig.~\ref{wavefunc1}(a). As a consequence, the infinite Hilbert space
is reduced to the product space of the open virtual bonds and the
remaining physical bonds in the cluster as
Fig.~\ref{wavefunc1}(b). The long range entanglement is gradually
considered by taking larger block sizes. Due to the modification of
the boundary sites of the cluster, an iPEPS wavefunction is converted
to an open boundary (OB) PEPS of size $l_x\times l_y$, of which the
contraction can be done efficiently at the wavefunction level. This
conversion is only made at the evolution stage. Once the ground state
tensors are obtained, one evaluates expectation values such as the
energy and magnetization using the Monte Carlo sampling
technique~\cite{ling} to a system with periodic boundary (PB)
condition and of system size $L\times L$.

The iterative steps of the cluster update are:
\begin{itemize}
\item write the iPEPS wavefunction into an OB PEPS using the
  $\mathbf{\Lambda}$s from the previous iterations as
  Fig.~\ref{wavefunc1}(a),
\item act with a local (imaginary time) evolution operator on several
  sites located in the center of the $l_x\times l_y$ cluster as
  Fig.~\ref{wavefunc1}(b),
\item contract every virtual bond except the one with the evolution
  operator, obtain the projectors and a new $\mathbf{\Lambda}$ as
  discussed below, and project the enlarged bond back to $D$,
\item finally replace the iPEPS wavefunction by the updated tensors as
  Fig.~\ref{wavefunc1}(c), and move to a different bond.
\end{itemize}

There are two main choices to be made in the above update procedure: first of all,
how to contract the environment tensor efficiently to two sites under
the time evolution; second, how to obtain the updated tensor.

\begin{figure}
\begin{center}
\includegraphics[width=8.cm]{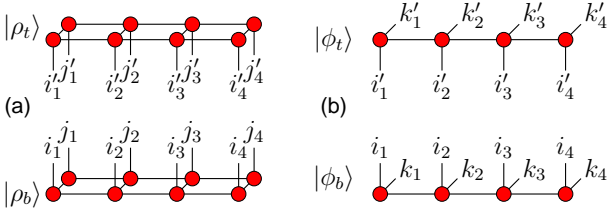}
\caption{(a) MPSs $|\rho_b\rangle$ and $|\rho_t\rangle$ of a
  conventional contraction method dealing with the norm
  $\langle\psi|\psi\rangle$ of a wavefunction $|\psi\rangle$, (b) MPSs
  $|\phi_b\rangle$ and $|\phi_t\rangle$ of a novel contraction method
  dealing with the wavefunction $|\psi\rangle$ directly.}
\label{mps-bt}
\end{center}
\end{figure}

To answer the first question, let us remind the conventional way of
contracting the environment tensor. In the conventional
evolution methods, one constructs a double tensor network, which
represents the norm $\langle \psi|\psi\rangle$ of the state
$|\psi\rangle$, then contracts this norm to obtain the environment
tensor of two sites. The major approximation is to represent the
bottom (top) rows of the norm by an MPS:
$|\rho_b(\bar{i},\bar{j})\rangle\equiv\mbox{Tr}_{\bar{k}}\langle\phi_b(\bar{i},\bar{k})|\phi_b(\bar{j},\bar{k})\rangle$
($|\rho_t(\bar{i}^{\prime},\bar{j}^{\prime})\rangle\equiv\mbox{Tr}_{\bar{k^{\prime}}}\langle\phi_t(\bar{i^{\prime}},\bar{k^{\prime}})|\phi_t(\bar{j^{\prime}},\bar{k^{\prime}})\rangle$)
as Fig.~\ref{mps-bt}(a), where $|\phi_b(\bar{i},\bar{k})\rangle$
($|\phi_t(\bar{i}^{\prime},\bar{k}^{\prime})\rangle$) is the
wavefunction of the bottom (top) rows with the highest (lowest)
vertical bonds open, as Fig.~\ref{mps-bt}(b),
$\bar{i}\equiv\{i_1,i_2,i_3,i_4\}$, etc. However, during the
contraction, one ignores the fact that
$|\rho_b(\bar{i},\bar{j})\rangle$ is hermitian if written as a density
matrix of the fictitious degrees of freedom $\bar{i}$ and $\bar{j}$,
and treats $|\rho_b(\bar{i},\bar{j})\rangle$ as a general MPS with
degrees of freedom $\{i_l,j_l\}$ at each site $l$. This will cause
ill-condition if the truncation error is large. We therefore contract
the wavefunction directly to ensure the hermiticity of the norm. The
degrees of freedom at each site $l$ of
$|\phi_b(\bar{i},\bar{k})\rangle$ is $\{i_l,k_l\}$ with $i_l$ represents
the topmost vertical index and $k_l$ represents the renormalized
physical index of the $l^{\mbox{th}}$ bottom-half-column. The argument
for $|\phi_b(\bar{i},\bar{k})\rangle$ equally applies to
$|\phi_t(\bar{i}^{\prime},\bar{k}^{\prime})\rangle$ throughout the
paper.

\begin{figure}
\begin{center}
\includegraphics[width=8.5cm]{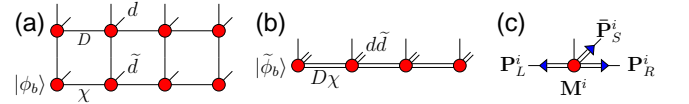}
\caption{(a) Renormalize a row into the MPS $|\phi_b\rangle$. (b)
  Absorb a row into $|\phi_b\rangle$ enlarges both the internal bond
  and the physical bond by a factor $D$ and $d$ respectively. (c)
  Projectors (blue triangles) $\mathbf{P}^i_L$, $\mathbf{P}^i_{R}$ and
  $\bar{\mathbf{P}}^i_S$ is calculated to reduce both the internal
  bond and physical bond of matrix $\mathbf{M}^i$.}
\label{mps-reduction}
\end{center}
\end{figure}

To contract a tensor network wavefunction with OB conditions, one
absorbs a row into $|\phi_b\rangle$ to obtain
$|\widetilde{\phi}_b\rangle$, as Fig.~\ref{mps-reduction}(a,b).
Unlike in the case of contracting the norm, this is described as a
matrix product operator (MPO) acting on an MPS, in such a way that not
only the internal bond dimension $\chi$ but also the physical degrees
of freedom $\tilde{d}$ of $|\phi_b\rangle$ is increased by a factor of
$D$ and $d$ respectively as Fig.~\ref{mps-reduction}(b), where $D$ and
$d$ are the virtual bond dimension and the physical degrees of freedom
in the original lattice. In order to keep the renormalization under
control, one has to truncate both, as Fig.~\ref{mps-reduction}(c).

\begin{figure}
\begin{center}
\includegraphics[width=8.cm]{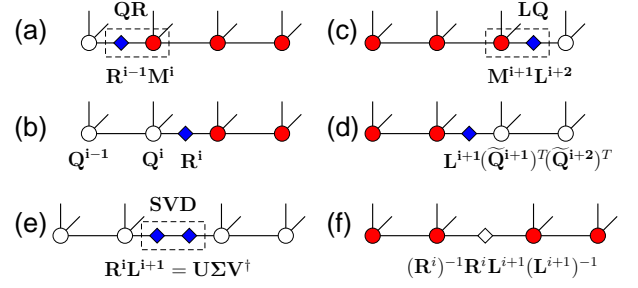}
\caption{(a-d) Calculate the residue matrix (blue diamonds)
  $\mathbf{R_i}$ ($\mathbf{L_i}$) of the left (right) half chain up to
  site $i$ by iteratively using QR (LQ) decompositions. (e) Replace
  left (right) half chain by $\mathbf{R}^i$ ($\mathbf{L}^{i+1}$) and
  use an SVD to reduce the dimension of the $i$-$\mbox{th}$ bond. (f)
  Inserted an identity (hollow diamond) at bond $i$ in original MPS
  and replace
  $\mathbf{R}^i\mathbf{L}^{i+1}\approx\bar{\mathbf{U}}\bar{\mathbf{\Sigma}}\bar{\mathbf{V}}^{\dagger}$
  to obtain the projectors $\mathbf{P}^i_R$ and $\mathbf{P}^{i+1}_L$.}
\label{projector_mps}
\end{center}
\end{figure}

\begin{figure}
\begin{center}
\includegraphics[width=8.5cm]{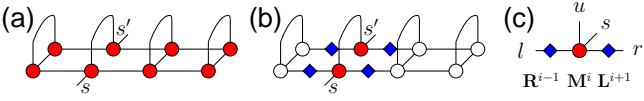}
\caption{(a) Reduced density matrix of the physical bond $s$ at site
  $i$ of $|\widetilde{\phi}_b\rangle$. (b,c) A reformation of (a) by
  $\mathbf{M}^i$ and the left and right residue matrices
  $\mathbf{L}^{i+1}$, $\mathbf{R}^{i-1}$.}
\label{reduce_physical}
\end{center}
\end{figure}

To reduce the bond dimension between site $i$ and $i+1$ of an OB MPS, a singular
value decomposition (SVD) is done on  $|\widetilde{\phi}_b\rangle$. As in
Fig.~\ref{mps-reduction}(e), one needs the right (left) residue matrix
$\mathbf{R}^i$ ($\mathbf{L}^{i+1}$) of the left (right) half chain up
to the site $i$ ($i+1$) to bring them into their normal
form~\cite{valentinadvphys}. To be specific, one calculates the
boundary residue matrix $\mathbf{R}^1$ ($\mathbf{L}^{n}$) using a QR
(LQ) decomposition, $n$ is the length of $|\widetilde{\phi}_b\rangle$,
\begin{eqnarray}
M^1_{usr}=\sum_{r^{\prime}}Q^1_{us,r^{\prime}}R^1_{r^{\prime},r},\\
M^n_{lus}=\sum_{l^{\prime}}L^n_{l,l^{\prime}}(\widetilde{Q}^{n}_{l^{\prime},us})^T,
\end{eqnarray}
then moves one site to the right (left) as in Fig.~\ref{projector_mps}(a-d),
\begin{eqnarray}
\sum_{l^{\prime}}R^{i-1}_{l,l^{\prime}}M^{i}_{l^{\prime},usr}=\sum_{r^{\prime}}Q^i_{lus,r^{\prime}}R^i_{r^{\prime},r},\\
\sum_{r^{\prime}}M^i_{lus,r^{\prime}}L^{i+1}_{r^{\prime},r}=\sum_{l^{\prime}}L^{i}_{l,l^{\prime}}(\widetilde{Q}^{i}_{l^{\prime},usr})^T,
\end{eqnarray}
where $\mathbf{M}^i$ is the matrix at the site $i$ of
$|\widetilde{\phi}_b\rangle$, $l,r,u,s$ represent the left, right, up
and physical index respectively, and $\mathbf{Q}^i$
($\widetilde{\mathbf{Q}}^{i}$) is isometric matrix. Thus the left
(right) half chain, which is replaced by $\mathbf{R}^i$
($\mathbf{L}^{i+1}$), is in its normal form as
Fig.~\ref{projector_mps}(e). An SVD
$\mathbf{R}^i\mathbf{L}^{i+1}\approx\bar{\mathbf{U}}\bar{\mathbf{\Sigma}}
\bar{\mathbf{V}}^{\dagger}$ will minimize
$||\widetilde{\phi}_b\rangle-|\phi_b^{\prime}\rangle||^2$, where
$|\phi_b^{\prime}\rangle$ is the one-bond reduced MPS to
$|\widetilde{\phi}_b\rangle$, bar means taking the leading singular
values or vectors. To derive the projectors for this reduction, one
inserts an identity
$(\mathbf{R}^{i})^{-1}\mathbf{R}^{i}\mathbf{L}^{i+1}(\mathbf{L}^{i+1})^{-1}$
at bond $i$ in the original MPS and substitutes
$\mathbf{R}^i\mathbf{L}^{i+1}$ by
$\bar{\mathbf{U}}\bar{\mathbf{\Sigma}} \bar{\mathbf{V}}^{\dagger}$ as
Fig.~\ref{projector_mps}(f). Equally distributing
$\bar{\mathbf{\Sigma}}$ to each side, one writes the projectors at
bond $i$ as
\begin{equation}
\mathbf{P}^i_R=(\mathbf{R}^{i})^{-1}\mathbf{\bar{U}}\sqrt{\mathbf{\bar{\Sigma}}},\quad
(\mathbf{P}^{i+1}_L)^{T}=\sqrt{\mathbf{\bar{\Sigma}}}\mathbf{\bar{V}}^{\dagger}(\mathbf{L}^{i+1})^{-1}.
\end{equation}
To avoid the matrix inversion, one can use
$(\mathbf{R}^i\mathbf{L}^{i+1})^{-1}=\mathbf{V}\frac{1}{\mathbf{\Sigma}}\mathbf{U}^{\dagger}$
to rewrite the projectors as
\begin{equation}
\label{pr}
\mathbf{P}^i_R=\mathbf{L}^{i+1}\mathbf{\bar{V}}\frac{1}{\sqrt{\mathbf{\bar{\Sigma}}}},\quad
(\mathbf{P}^{i+1}_L)^{T}=\frac{1}{\sqrt{\mathbf{\bar{\Sigma}}}}\mathbf{\bar{U}}^{\dagger}\mathbf{R}^i;
\end{equation}
note that taking $\mathbf{\bar{\Sigma}}^{-1}$ will not cause a
singularity as the small singular values are discarded. To reduce the
physical degrees of freedom of $\mathbf{M}^i$, one considers the
reduced density matrix $\rho^i_{s,s^{\prime}}$ of site $i$ as
Fig.~\ref{reduce_physical}. Define
$\widetilde{M}^i_{lusr}=\sum_{l^{\prime}r^{\prime}}R^{i-1}_{l,l^{\prime}}M^i_{l^{\prime}usr^{\prime}}L^{i+1}_{r^{\prime},r}$,
then
$\rho^i_{s,s^{\prime}}=\sum_{lur}\widetilde{M}^{i}_{s,lur}\widetilde{M}^{i\dagger}_{lur,s^{\prime}}$. The
leading eigenvectors $\bar{\mathbf{P}}^i_S$ of
$\rho^i=\mathbf{P}^{i}_S\mathbf{\Lambda}^i\mathbf{P}^{i\dagger}_S$ is
the projector to reduce the physical degrees of freedom of
$\mathbf{M}^i$. Upon obtaining $\mathbf{P}^i_L$, $\mathbf{P}^i_R$ and
$\bar{\mathbf{P}}^i_S$, one projects according to
Fig.~\ref{mps-reduction}(c). Note that one left (right) sweep can
obtain $\mathbf{R}^i$ ($\mathbf{L}^i$) for all sites. It turns out
that the reduction of each internal bond and physical bond can be done
simultaneously as if they are independent. This concludes the question
of the contraction at the single-layer wavefunction level.

\begin{figure}
\begin{center}
\includegraphics[width=8cm]{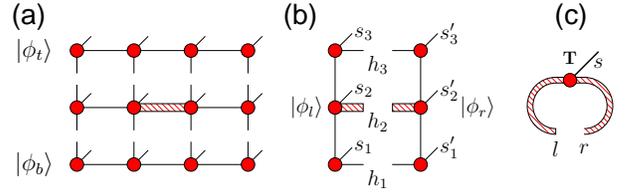}
\caption{(a) Bottom and top MPS $|\phi_b\rangle$, $|\phi_t\rangle$
  sandwich a row with the evolution operator. (b) Further absorb all
  columns into the left (right) MPS $|\phi_l\rangle$
  ($|\phi_r\rangle$). (c) Contract all virtual bonds in (b) except
  $h_2$ to form a single tensor $\mathbf{T}$.}
\label{contraction}
\end{center}
\end{figure}

Now we focus on the question of how to obtain the evolved tensors using the
partially contracted wavefunction. After renormalizing rows from below
(above) into $|\phi_b\rangle$ ($|\phi_t\rangle$) as
Fig.~\ref{contraction}(a), one renormalizes columns from left (right)
into $|\phi_l(\bar{h},\bar{s}_l)\rangle$
($|\phi_r(\bar{h},\bar{s}_r)\rangle$), as in Fig.~\ref{contraction}(b),
where the thickened bond is due to the action of the evolution
operator. Define
\begin{equation}
\label{phi}
|\widetilde{\psi}(\bar{s}_l,\bar{s}_r)\rangle=\sum_{\bar{h}}|\phi_l(\bar{h},\bar{s}_l)\rangle\otimes\phi_r(\bar{h},\bar{s}_r)\rangle,
\end{equation}
where $\bar{h}=\{h_1,h_2,h_3\}$ is the horizontal internal bond, the
projector to reduce the thickened bond has to minimize the 2-norm
$||\widetilde{\psi}(\bar{s}_l,\bar{s}_r)\rangle-|\psi^{\prime}(\bar{s}_l,\bar{s}_r)\rangle||^2$.
Unlike in the OB MPS, where one cut will bipartite the wavefunction,
here one deals with a ring MPS. We present an empirical way to calculate the projectors.
First contract all virtual bonds in
$|\widetilde{\psi}(\bar{s}_l,\bar{s}_r)\rangle$ except the thickened
bond $h_2$ to form a single tensor $\mathbf{T}$ as
Fig.~\ref{contraction}(c); in analogy to an OB MPS, make a
QR (LQ) decomposition to calculate the right (left) residue matrix
$\mathbf{R}$ ($\mathbf{L}$) of the tensor $\mathbf{T}$ as
\begin{eqnarray}
T_{rs,l}=\sum_{l^{\prime}}Q_{rs,l^{\prime}}R_{l^{\prime},l},\\
T_{r,sl}=\sum_{r^{\prime}}L_{r,r^{\prime}}\widetilde{Q}^{T}_{r^{\prime},sl},
\end{eqnarray}
where $\mathbf{Q}$ ($\widetilde{\mathbf{Q}}$) is isometric matrix;
insert $\mathbf{R}^{-1}\mathbf{RLL}^{-1}$ between the $l,r$ indices of
tensor $\mathbf{T}$, in analogy to Eq.~\ref{pr}, derive the projectors
to the imaginary time evolution as
\begin{equation}
\mathbf{P}_A=\mathbf{L\bar{V}}\frac{1}{\sqrt{\mathbf{\bar{\Lambda}}}},\quad
(\mathbf{P}_B)^{T}=\frac{1}{\sqrt{\mathbf{\bar{\Lambda}}}}\mathbf{\bar{U}^{\dagger}R},
\end{equation}
where an SVD
$\mathbf{RL}\approx\bar{\mathbf{U}}\bar{\mathbf{\Lambda}}\bar{\mathbf{V}^{\dagger}}$
is performed.  $\bar{\mathbf{\Lambda}}$ for this particular bond
contains the entanglement spectrum that replaces the previous one for
the next evolution step, analogous to what happens in the simple
update scheme. Calculating the $\mathbf{T}$ tensor explicitly is
expensive, what one does instead when dealing with the object
Fig.~\ref{contraction}(c) is treating the $r$ index as the physical
ones $\{\bar{s}_l,\bar{s}_r\}$ to calculate the $\mathbf{R}$ matrix
and again treating the $l$ index as the physical ones to calculate the
$\mathbf{L}$ matrix. Another way to update the projectors
$\mathbf{P}_A$, $\mathbf{P}_B$ would be alternatively solving the
multi-quadratic equations~\cite{valentinadvphys}, where the
$\mathbf{T}$ tensor is explicitly needed.

\begin{figure}
\begin{center}
\includegraphics[width=6.5cm]{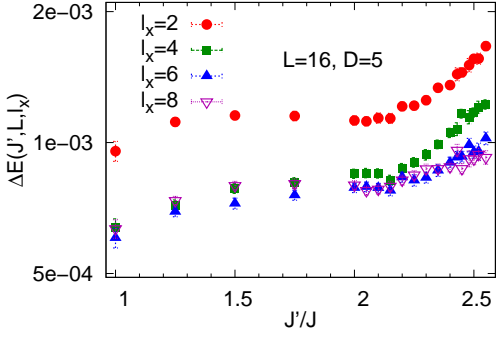}
\caption{(a) Ground state energy errors per site for $L=16$ using tensors of
  bond dimension $D=5$ obtained via the simple update ($l_x=2$) and
  the cluster update ($l_x=4,6,8$).}
\label{err16}
\end{center}
\end{figure}

\begin{figure}
\begin{center}
\includegraphics[width=6.5cm]{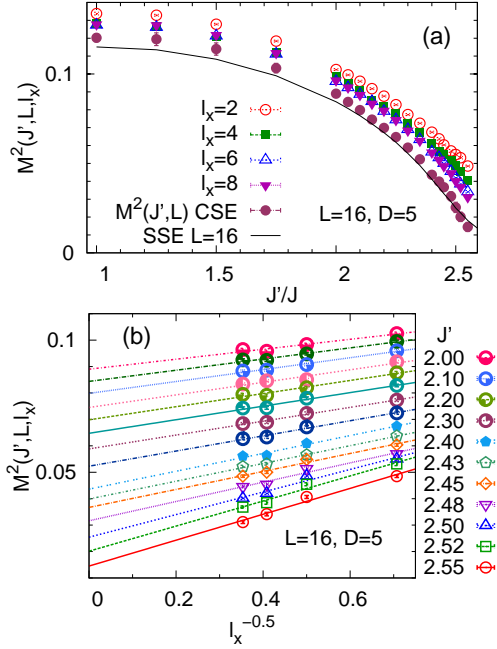}
\caption{(a) Sublattice magnetization square for $L=16$ using the same
  parameters as Fig.~\ref{err16}, black solid line is the SSE results
  for $L=16$, solid circles are the CSE results from (b). (b) The
  cluster size extrapolations (CSE) for the finite size ($L=16$)
  magnetizations using cluster sizes $l_x=2,4,6,8$.}
\label{mag16}
\end{center}
\end{figure}

{\it Results} -- We benchmark our method using the spin 1/2 staggered
dimerized antiferromagnetic Heisenberg model on square lattice with PB
conditions and compare our results to that obtained from the simple
update~\cite{linprb84}, iPEPS method~\cite{bauerjstatmech} and the QMC
simulation~\cite{wenzel}. The Hamiltonian is written as
\begin{equation}
H=J\sum_{\langle i,j\rangle}\mathbf{S}_i\cdot\mathbf{S}_j+J^{\prime}\sum_{\langle\langle k,l\rangle\rangle}\mathbf{S}_k\cdot\mathbf{S}_l,\quad (J,J^{\prime}>0),
\end{equation}
where $\langle\langle k,l\rangle\rangle$ are the horizontal nearest
neighbor pairs satisfying $\mbox{mod}(x_k,2)=\mbox{mod}(y_k,2)$,
$x_k,y_k$ is the lattice coordinate of site $k$, $\langle i,j\rangle$
are all other nearest neighbor pairs, and $J,J^{\prime}$ are the
coupling strengths. This model is frustration free and has been
extensively studied~\cite{wenzel} using the stochastic series
expansion (SSE) method~\cite{sandvikprb56-11678}. Increasing the
coupling strength ratio $J^{\prime}/J$ drives the system through a
second order phase transition from the antiferromagnetic ordered phase
to a magnetically disordered phase. We set $J=1$ for convenience
hereafter.

We obtain the finite size ground state energies and the magnetization
for systems with linear size $L=4,8,16$. Each system is measured using
the variational monte carlo (VMC) sampling technique~\cite{ling}
taking the tensors of bond dimension $D=5$ obtained through the
cluster update with cluster size $l_x\times l_y=2\times 1$, $4\times
4$, $6\times 6$ and $8\times 8$. For this specific model, the simple
update is equivalent to a cluster update with cluster size $2\times
1$.

\begin{figure}
\begin{center}
\includegraphics[width=6.5cm]{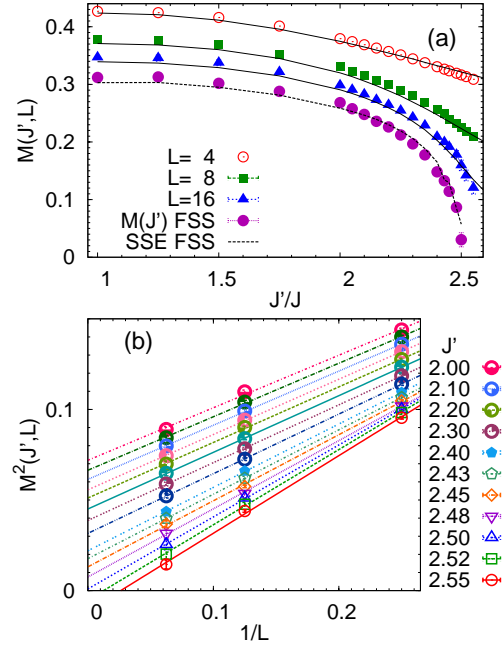}
\caption{(a) Finite size magnetizations and their thermodynamic limits
  via the FSS from (b), the black solid lines are the SSE simulation
  results for system size $L=4,8,16$ and the black dash line is their
  FFS results. (b) Finite size scaling of the magnetization square
  using system sizes $L=4,8,16$.}
\label{magallsize}
\end{center}
\end{figure}

\begin{figure}
\begin{center}
\includegraphics[width=6.cm]{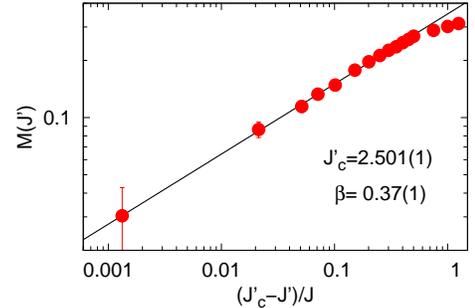}
\caption{ Fitting of the magnetization curve for the critical value
  $J^{\prime}_c$ and the critical exponent $\beta$.}
\label{order}
\end{center}
\end{figure}

We present the ground state energy errors per site for system size
$L=16$ with $D=5$ in Fig.~\ref{err16} and the sublattice magnetization
square defined as following~\cite{sandvikprb56-11678} in
Fig.~\ref{mag16},
\begin{equation}
M^2=\frac{1}{L^2}\sum_i\mathbf{S}_{x_i,y_i}\cdot\mathbf{S}_{x_i+\frac{L}{2},y_i+\frac{L}{2}}.
\end{equation}
Significant improvement in energy and magnetization is achieved by
increasing the cluster size $l_x$ from 2 to 8. We make a cluster size
extrapolation (CSE) for the finite size magnetization in
Fig.~\ref{mag16}(b) as
\begin{equation}
M^2(J^{\prime},L,l_x)=M^2(J^{\prime},L)+a_{J^{\prime}}/\sqrt{l_x},
\end{equation}
where $M^2(J^{\prime},L)$ and $a_{J^{\prime}}$ are the fitting
parameters. The CSE results are in good agreement with the finite size
magnetization through the SSE simulation, as plotted in
Fig.~\ref{mag16}(a). Here the cluster size $l_x$ plays the role of the
bond dimension $D$ to provide a scaling scheme. To obtain the
sublattice magnetization in the thermodynamic limit, we use the
following finite size scaling (FSS)
formula~\cite{neubergerprb39,fisherprb39,sandvikprb56-11678} to
extrapolate $M(J^{\prime})$ as in Fig.~\ref{magallsize}(b)
\begin{equation}
\label{fss}
M^2(J^{\prime},L)=M^2(J^{\prime})+b_{J^{\prime}}/L,
\end{equation}
where $M^2(J^{\prime})$ and $b_{J^{\prime}}$ are the fitting
parameters. This FSS formula is only valid for the states with
persisting sublattice magnetization, thus the extrapolated negative
$M^2(J^{\prime})$ simply means that these states have no magnetic
order. The finite size magnetizations $M(J^{\prime},L)$ together with
their thermodynamic limits $M(J^{\prime})$ are plotted in
Fig.~\ref{magallsize}(a). The extrapolated $M(J^{\prime})$ lies a bit
off the SSE FSS results near the critical point, because the latter are
fitted with a sub-leading corrections $1/L^2$ using system sizes
$L=4,8,16,32$. To determine the critical value $J^{\prime}_c$ and the
critical exponent $\beta$, we fit $M(J^{\prime})$ by
\begin{equation}
M(J^{\prime})=c (J_c^{\prime}-J^{\prime})^{\beta},
\end{equation}
where $J^{\prime}_c$, $\beta$ and $c$ are the fitting parameters, as
in Fig.~\ref{order}. We obtain a critical value
$J^{\prime}_c=2.501(1)$ and the critical exponent $\beta=0.37(1)$.
The SSE results are $J^{\prime}_c=2.5198(3)$ and
$\beta=0.376(5)$~\cite{wenzel}. The offset in $J^{\prime}_c$ is due to
ignoring the sub-leading corrections in the FSS formula Eq.~\ref{fss}.
Considering it would require results from a much larger bond dimension
$D$ for system size $L=32$. The critical value from the simple update
is $J_c^{\prime}=2.56$~\cite{linprb84}, and from the iPEPS is
$J_c^{\prime}\approx 2.85$~\cite{bauerjstatmech}.

{\it discussion} -- The cluster update allows one to accurately
determine the behavior of a second order phase transition using a
tensor network state of intermediate bond dimension $D$ and relatively
small cluster size $l_x$. The simple update is efficient, however it
always generates a fat tail near the exact transition point; on the
contrary, the cluster update that has been introduced here improves
those results significantly. This improvement is especially important
when a narrow intermediate phase is present in the phase diagram, such
as in the frustrated $J_1-J_2$ Heisenberg model on square
lattice. Comparing to the complete contraction algorithm in
~\cite{iztok}, the Hilbert space is drastically reduced, which
significantly boosts the efficiency, and the algorithm is simpler in
the sense that it is free of dealing with the boundary evolutions. The
cluster update scales as $D^{4}\chi^2d\tilde{d}$. The renormalized
physical index $\tilde{d}$ varies depending on the entanglement of the
state, however it is bounded by $D\chi^2$. Choosing a small cluster
size also relaxes the scaling of $\tilde{d}$, {\it e.g.} the
complexity with cluster size $2\times 2$ scales as $D^5$, and $4\times
4$ scales as $D^7$ if taking $\chi=D$.

{\it Conclusion} -- We presented a cluster imaginary time evolution
method for a tensor network state (TNS) describing the ground state of
strongly correlated quantum systems. We benchmarked this method
with the staggered dimerized antiferromagnetic Heisenberg model on the
square lattice and accurately determined its critical value and
critical exponents $\beta$ using a TNS with fairly small bond
dimension $D=5$; this provides clear evidence for an improvement over the simple
update scheme in ~\cite{xiang08}. The efficiency and accuracy of
this method should allow tensor network simulations to be applied to a zoo of
interesting models that are not easily accessible by other methods, especially as large values of $D$ can be treated \cite{progress}.

{\it Acknowledgments:} This project is supported by the EU Strep
project QUEVADIS, the ERC grant QUERG, and the FWF SFB grants FoQuS
and ViCoM. The computational results presented have been achieved
using the Vienna Scientific Cluster (VSC).

\end{document}